\documentclass[12pt]{article}
\usepackage{german}
\usepackage{jb}

\newcommand{\eqd}[1]{#1}
\newcommand{\eqr}[1]{#1}

\newcommand{\eqS}[1]{}

\newcommand{\V}{{\cal V}}
\newcommand{\CR}{{\cal CR}}
\newcommand{\T}{{\cal T}}

\renewcommand{\P}{{\cal P}}

\newcommand{\vx}{{\vec{x}}}

\newcommand{\eps}{\varepsilon}
\renewcommand{\phi}{\varphi}

\newcommand{\pref}{\mathrel{\mathchoice{\;\mbox{\sf pref}\;}
				{\;\mbox{\sf pref}\;}
				{\;\mbox{\scriptsize\sf pref}\;}
				{\;\mbox{\scriptsize\sf pref}\;} }}

\newcommand{\npref}{\mathrel{\;\mbox{\sf pr/\hspace{-0.5em}ef}\;}}
\newcommand{\sibl}{\mathrel{\mathchoice{\;\mbox{\sf sibl}\;}
				{\;\mbox{\sf sibl}\;}
				{\;\mbox{\scriptsize\sf sibl}\;}
				{\;\mbox{\scriptsize\sf sibl}\;} }}
\newcommand{\at}[1]{/_{#1}}
\newcommand{\equivs}{\stackrel{*}{\equiv}}
\newcommand{\precs}{\stackrel{*}{\prec}}
\newcommand{\preceqs}{\stackrel{*}{\preceq}}
\newcommand{\pathps}{\mathop{paths}}

\newcommand{\display}[2]{
	\par
	{\bf \optional{}{#1}{:}}
	\ 
	\hspace*{\fill}
	#2
	\hspace*{\fill}
	\ 
	\par
	}

\newcounter{lit}

\newcommand{\N}{I\!\!N}
\newcommand{\tpl}[1]{\langle#1\rangle}
\newcommand{\la}{\leftarrow}
\newcommand{\ra}{\rightarrow}
\newcommand{\Ra}{\Rightarrow}
\newcommand{\Lra}{\Leftrightarrow}

\newcommand{\mca}[2]{\multicolumn{#1}{@{}c@{}}{#2}}
\renewcommand{\.}[1]{\!#1\!}

\newcommand{\dfn}[2]{%
	\par%
	{\bf Definition #1}
	~
	#2%
	\par%
}

\newcommand{\alg}[2]{%
	\par%
	{\bf Algorithmus #1}
	~
	#2%
	\par%
}

\newcommand{\bsp}[2]{%
	\par%
	{\bf Beispiel #1}
	~
	#2%
	\par%
}

\newcommand{\lemmal}[2]{%
	\par%
	{\bf Lemma #1}
	~
	#2%
	\par%
}

\newcommand{\satZ}[2]{%
	\par%
	{\bf Satz #1}
	~
	#2%
	\par%
}

\newcommand{\algl}[2]{\alg{#1}{#2}}

\newcommand{\be}{\begin{enumerate}}
\newcommand{\ee}{\end{enumerate}}

\newcommand{\bi}{\begin{itemize}}
\newcommand{\ei}{\end{itemize}}

\newcommand{\bd}{\begin{description}}
\newcommand{\ed}{\end{description}}

\newcommand{\tab}{\hspace*{1cm}}

\newcommand{\dt}[1]{%
	\refstepcounter{lit}%
	\item[{\rm [\arabic{lit}]}]%
	\label{#1}}
\newcommand{\cit}[1]{[\ref{#1}]}

\begin{document}


\begin{center}
{\Large\bf
Eine entscheidbare Klasse $n$-stelliger Horn-Pr\"adikate

}
\vspace*{0.2cm}

Jochen Burghardt
\vspace*{0.2cm}

GMD,
Rudower Chaussee 5,
D-12489 Berlin
\\
Tel   *49-30-6392-1867,
Fax   -1805,
E-Mail jochen@first.gmd.de
\vspace*{1cm}

\end{center}

\subsection{Einleitung}
\label{Einleitung}

Das Rechnen mit unendlichen Mengen von Grundtermen hat zahlreiche
Anwendungen, speziell auf den Gebieten der Programmanalyse und des
automatischen Beweisens:
z.B.\
Typ-Inferenz in PROLOG, basierend auf einer oberen Absch\"atzung der
Extension von Pr\"adikaten \cit{Mishra84};
Berechnung einfacher Invarianten imperativer Programme \cit{Jones79};
Beschreibung von Bildmengen von Abstraktionsfunktionen in
Implementierungsbeweisen \cit{Burghardt93};
Beschreibung der \"Aquivalenzklassen von Grundtermen f\"ur
E-Anti-Unifikation \cit{Heinz94}; usw.

Allen Anwendungen ist als Mindestanforderung gemeinsam, da\3 der
Durchschnitt zweier darstellbarer Termmengen wieder darstellbar sein
mu\3 und da\3 entscheidbar sein mu\3, ob eine dargestellte Menge leer
ist.
In diesem speziellen Sinne wird hier von Entscheidbarkeit gesprochen.
Die Leistungsf\"ahigkeit eines Formalismus zur Beschreibung solcher
Termmengen h\"angt entscheidend ab von der Komplexit\"at der Mengen,
die er darzustellen in der Lage ist.

Der bekannteste Beschreibungsformalismus, regul\"are Baumautomaten
\cit{Comon90}, ist
\"aqui\-va\-lent zu Hornklauseln mit einstelligen Pr\"adikaten und
ausschlie\3lich Variablen als Argumenten im Rumpf, bzw.\ in der
Terminologie von \cit{Schmidt-Schauss88} zu linearen Termdeklarationen.

In \cit{Uribe92} werden semi-lineare Termdeklarationen zugelassen,
d.h.\ mehrfache Vorkommen einer Variablen werden
erlaubt, sofern die Listen
der Funktionssymbole auf den Pfaden von der Wurzel zu den Vorkommen
jeweils gleich sind;
\"aquivalent dazu sind regul\"are Baumautomaten mit Gleichheitstests
f\"ur direkte Unterterme \cit{Bogaert92},
f\"ur nicht-direkte Unterterme wird die Disjunktheit der Sprachen
bereits unentscheidbar \cit{Tommasi91}.

In \cit{Burghardt93} werden regul\"are Substitutionsmengen zugelassen,
\"aquivalent dazu sind $n$-stellige Horn-Pr\"adikate mit linearen
Kopftermen $f_1(\vx_1),\ldots,f_m(\vx_m)$ und linearen
Variablenvektoren $\vx^1,\ldots,\vx^n$ als Rumpfargumenten, sofern die
$\vx_i$ die Zeilen- und die $\vx^j$ die Spaltenvektoren einer
verallgemeinerten Matrix bilden.
Damit sind auch einfache Relationen wie z.B.\
$x \leq y$ f\"ur $x,y \in \N$
ausdr\"uckbar.

$\Omega$-Terme \cit{Comon92} entsprechen
speziellen Hornklauselmengen, in denen f\"ur jedes Pr\"adikat genau
zwei Klauseln existieren, quasi eine f\"ur den Basis- und eine f\"ur
den Rekursionsfall.
Durch die starke Einschr\"ankung der m\"oglichen Rekursion auf die
Struktur von $\N$ wird es m\"oglich, da\3 eine beschr\"ankte Anzahl von
Variablen beliebig oft und an beliebigen Stellen
in beschreibenden $\Omega$-Termen vorkommen darf.

Im folgenden wird ein Ansatz vorgestellt, der auf einer mit der
Unifikation kommutierenden Termordnung basiert
und teilweise
ausdrucksst\"arker ist als semi-lineare Term-Deklarationen.

\subsection{\"Uberblick}
\label{Uberblick}

Sei eine endliche Menge $\CR$ von Funktionssymbolen 
(genauer: Konstruktorsymbolen) mit ihren festen
Stelligkeiten und eine
unendliche Menge $\V$ von Variablen gegeben;
sei $\T$ die Menge aller Terme \"uber $\V \cup \CR$.
Terme werden mit $t, t', t_i, \ldots$ bezeichnet,
Variablen mit $x, y, z, \ldots$,
Funktionssymbole mit $f, g, \ldots$,
Pr\"adikate mit $p, q, \ldots$,
Substitutionen mit $\beta, \gamma, \ldots$,
$\tpl{t_1,\ldots,t_n}$ bezeichnet ein $n$-Tupel von Termen,
$vars(t)$ bezeichnet die Menge aller in $t$ vorkommenden Variablen,
$dom(\beta) := \{ x \in \V \mid \beta x \neq x\}$ bezeichnet den
Definitionsbereich einer Substitution.
Substitutionen werden oBdA.\ stets als idempotent und
mit gen\"ugend gro\3em Definitionsbereich angenommen,
so da\3 f\"ur alle vorkommenden Anwendungen $\beta t$
stets $vars(t) \subset dom(\beta)$ ist.
Bezeichnet $\beta = mgu(t_1,t_2)$ die unifizierende
Substitution f\"ur $t_1$ und $t_2$, falls sie
existiert, 
so ist $lci(t_1,t_2) = \beta t_1 = \beta t_2$
die allgemeinste gemeinsame Instanz.
Zu einer gegebenen irreflexiven Ordnungsrelation $<$ bezeichnen wir mit
$\leq$ deren reflexive H\"ulle.

Eine Termmenge wird dargestellt als die Extension eines Pr\"adikats
$p$,
das durch ein Horn-Programm der Form
	\display{\eqd{1}}{
	\begin{tabular}{@{}l@{$\;$}l@{$\;$}l@{$\;$}l@{}}
	$p_1(t_1)$ & $\la p_{11}(t_{11})$ & $\wedge \ldots \wedge$
		& $p_{1n_1}(t_{1n_1})$	\\
		& \ldots	\\
	$p_m(t_m)$ & $\la p_{m1}(t_{m1})$ & $\wedge \ldots \wedge$
		& $p_{mn_m}(t_{mn_m})$	\\
	\end{tabular}
	}
gegeben ist, wobei die $p_i$, die $p_{ij}$ und $p$ keineswegs
verschieden sein m\"ussen.
Fakten werden durch Klauseln mit $n_i = 0$ dargestellt.
Die $t_i$ und die $t_{ij}$ sind beliebige Terme, die folgenden
Anforderungen gen\"ugen:

\dfn{\eqd{2}}{
\be
\item $vars(t_{ij}) \cap vars(t_{ij'}) = \{\}$ 
	f\"ur $1 \leq j \neq j' \leq n_i$,
\item $t_{ij} \preceqs t_i$ f\"ur $1 \leq i \leq m$
	und $1 \leq j \leq n_i$, 
\item dabei sei $\preceqs$ die reflexive H\"ulle einer Wohlordnung 
	$\precs$
	mit endlichem Verzweigungsgrad, die in folgendem Sinne mit
	der Unifikation kommutiert:
\item zu $t' \preceqs t = lci(t_1,t_2)$
	existieren stets $t'_1 \preceqs t_1$ und $t'_2 \preceqs t_2$
	mit $t' = lci(t'_1,t'_2)$,
\item $\preceqs$ sei abgeschlossen gegen alle Substitutionen $\beta$,
	die im Laufe der Algorithmen
	auftreten k\"onnen\footnote{
	Diese Einschr\"ankung wird sp\"ater formal pr\"azisiert, vgl.\
	Definition \eqr{14} ff.
	},
	d.h.\ $t' \preceqs t \Ra \beta t' \preceqs \beta t$.
\ee
}

Man beachte, da\3 $\preceqs$ nicht gegen beliebige Substitutionen
 abgeschlossen sein mu\3.
Die Pr\"adikate werden der Einfachheit halber als einstellig
geschrieben, Mehrstelligkeit l\"a\3t sich leicht durch $n$-Tupel
codieren;
in Beispielen verwenden wir dazu
``$:$'' als zweistelliges Funktionssymbol
niedrigster Priorit\"at.
Verlangt man, da\3 die $t_{ij}$ s\"amtlich Variablen sein m\"ussen und
die $t_i$ linear bzw.\ semi-linear, so erh\"alt man regul\"are
Baumsprachen bzw.\
semi-lineare Termdeklarationen.

Im folgenden werden die beiden Algorithmen $inh$ zur Entscheidung der
Erf\"ullbarkeit eines Pr\"adikats und ${\it inf}$
zur Berechnung der Konjunktion
zweier Pr\"adikate vorgestellt.
Beide verlangen z.Zt.\ noch die --~starke~--
Einschr\"ankung $n_i \leq 1$ f\"ur $i=1,\ldots,m$.

\alg{\eqd{3}}{(Erf\"ullbarkeit: $inh$)	\\
Sei ein Horn-Programm wie in \eqr{1} gegeben,
wobei als zus\"atzliche Einschr\"ankung
$n_i \leq 1$ f\"ur $i=1,\ldots,m$ sei.
$inh_\vee(p^{t'},\{\})$ liefert $true$ 
genau dann, wenn das Pr\"adikat $p$ durch eine
Instanz des Terms $t'$ erf\"ullbar ist.

\begin{tabular}[t]{@{}ll@{$\;$}l@{$\;$}ll@{}l@{}}
1. & $inh_\vee(p^{t'},Occ)$ & $\Lra inh_\wedge(p_1^{\beta_1 t_1},Occ)$
	& $\vee \ldots \vee 
	inh_\wedge(p_k^{\beta_k t_k},Occ)$ && $(*1)$	\\
2. & $inh_\wedge(p^{t'},Occ)$ & $\Lra false$ 
	&& f\"ur $p^{t'} \in Occ$,\hspace*{0.15cm} & $(*2)$	\\
3. & $inh_\wedge(p_i^{\beta_i t_i},Occ)$ 
	& \mca{2}{$\Lra inh_\vee(p_{i1}^{\beta_i t_{i1}},
	Occ \cup \{p_i^{\beta_i t_i}\})$}
	& f\"ur $n_i = 1$,	\\
4. & $inh_\wedge(p_i^{\beta_i t_i},Occ)$ & $\Lra true$
	&& f\"ur $n_i = 0$ \\
\end{tabular}

\begin{tabular}[t]{@{}l@{$\;$}ll@{}}
$(*1)$: & Dabei seien oBdA.\ $p_1 = \ldots = p_k = p$ 
	s\"amtliche Klauselk\"opfe f\"ur $p$,	\\
	& f\"ur die $\beta_i := mgu(t',t_i)$ existiert.	\\
$(*2)$: & Der Test mu\3 mod.\ gebundener Umbenennung 
	durchgef\"uhrt werden.	\\
\end{tabular}
}

\algl{\eqd{4}}{(Konjunktion: ${\it inf}$)	\\
Sei ein Horn-Programm wie in \eqr{1} gegeben,
mit den zus\"atzlichen Einschr\"ankungen
\bi
\item $n_i \leq 1$ f\"ur $i=1,\ldots,m$ und
\item $\tpl{t_{i1},t_{i'1}} \preceqs \tpl{t_i,t_{i'}}$
	f\"ur alle $i,i' \in \{1,\ldots,m\}$
	\tab
	(als Versch\"arfung von \eqr{2}.2).
\ei

Seien zwei Pr\"adikate $p_{i_1},p_{i_2}$
gegeben.
Der Algorithmus f\"uhrt einen neuen Pr\"a\-di\-kat\-na\-men ein,
der hier mit $p_{i_1} p_{i_2}$ bezeichnet wird,
und definiert neue Klauseln daf\"ur, 
so da\3 
$p_{i_1} p_{i_2}(x) \Lra p_{i_1}(x) \wedge p_{i_2}(x)$ 
wird:
\\
F\"ur alle Klauseln\footnote{
	OBdA.\ werden hier nur
	die Klauseln ${i_1}$ und ${i_2}$ selbst betrachtet.
	}
von $p_{i_1}$ und $p_{i_2}$,
f\"ur die $t_{i_1}$ und $t_{i_2}$
etwa durch $\beta$ unifizierbar sind,
definiere eine neue Klausel
$p_{i_1} p_{i_2}(\beta t_{i_1})
\la p_{{i_1}1} p_{{i_2}1}(\beta t_{{i_1}1})$.
\\
Verfahre in der gleichen Weise 
mit $p_{{i_1}1} p_{{i_2}1}$, usw.
}

Die partielle Korrektheit der beiden Algorithmen
ist jeweils relativ leicht zu zeigen, indem
die in \cit{Burghardt93} f\"ur regul\"are
Baumsprachen angewandte Methodik
entsprechend verallgemeinert wird; statt der Termstruktur
dient jetzt die Wohlordnung $\precs$ als Grundlage f\"ur die
Induktionsbeweise,
Einzelheiten werden in \cit{Burghardt95} ausgef\"uhrt.
Ohne die Einschr\"ankung $n_i \leq 1$
m\"u\3te Regel \eqr{3}.3 lauten
	\display{}{
	$inh_\wedge(p_i^{\beta_i t_i},Occ) 
	\Lra inh_\vee(p_{i1}^{\beta_i t_{i1}},Occ')
	\wedge \ldots \wedge
	inh_\vee(p_{in_i}^{\beta_i t_{in_i}},Occ')$
	f\"ur $n_i > 0$
	}
und es m\"u\3te sichergestellt werden, da\3 $\beta_i t_{i1}$, \ldots,
$\beta_i t_{in_i}$ paarweise disjunkte Variablen haben.
Z.Zt.\ wird untersucht, unter welchen Voraussetzungen
die Bedingung $n_i \leq 1$ fallengelassen
werden kann.
Damit k\"onnte der vorgestellte Ansatz u.U.\ eine echte Obermenge der
semi-linearen Termdeklarationen abdecken.

F\"ur die Terminierungsbeweise definieren wir
f\"ur $T \subset \T$:
\be
\item $lci(T) := \{ lci(t_1,\ldots,t_n) \mid t_1,\ldots,t_n \in T,
	lci(t_1,\ldots,t_n) \mbox{ existiert} \}$,
\item $less(T) := \{ t \mid t \preceqs t' \in T \}$,
\ee

F\"ur $T$ endlich ist $lci(T)$ und wegen \eqr{2}.3 auch
$less(T)$ endlich, also auch $lci \circ less(T)$.
Aus \eqr{2}.4 folgt sofort, da\3 
$less \circ lci(T) \subset lci \circ less(T)$.
Damit gilt:

\lemmal{\eqd{5}}{
Wird $inh_\vee(p^{t'},Occ)$ oder $inh_\wedge(p^{t'},Occ)$
direkt oder indirekt von $inh_\vee(q^{t''},\{\})$ aufgerufen,
so gilt
$t' \in lci \circ less(\{t_1,\ldots,t_m,t''\})$.
}{
Induktion \"uber die Indirektionsstufe des aktuellen Aufrufs:
F\"ur den ersten Aufruf von $inh$ ist nichts zu zeigen.
Fallunterscheidung, durch welche Regel des Algorithmus \eqr{3}
der aktuelle Aufruf entstanden ist:
\be
\item[1.]
	\begin{tabular}[t]{@{}l@{$\;$}ll@{}}
	& $\beta_i t_i$	\\
	$=$ & $lci(t',t_i)$ & nach $(*1)$ zu \eqr{3}.1	\\
	$\in$ & $lci \circ lci \circ less(\{t_1,\ldots,t_m,t''\})$
		& nach Induktionsvoraussetzung	\\
	$=$ & $lci \circ less(\{t_1,\ldots,t_m,t''\})$
		& da $lci$ idempotent	\\
	\end{tabular}
\item[2.,4.] Es entsteht kein neuer Aufruf.
\item[3.]
	\begin{tabular}[t]{@{}l@{$\;$}ll@{}}
	& $\beta_i t_{i1}$	\\
	$\in$ & $less \circ lci \circ less(\{t_1,\ldots,t_m,t''\})$
		& nach I.V., da 
		$\beta_i t_{i1} 
		\preceqs \beta_i t_i$	\\
	$\subset$ & $lci \circ less \circ less(\{t_1,\ldots,t_m,t''\})$
		& nach \eqr{2}.4, s.o.	\\
	$=$ & $lci \circ less(\{t_1,\ldots,t_m,t''\})$
		& da $less$ idempotent	\\
	\end{tabular}
\ee
}

Also treten nur endlich viele verschiedene ``Exponenten'' auf;
wegen der Regel \eqr{3}.2 wird $inh$ nie zweimal mit demselben
Argument (mod.\ Umbenennung) aufgerufen, so da\3 der
Algorithmus terminiert.
Die Terminierung von ${\it inf}$ ist trivial.

\bsp{\eqd{6}}{
$0$ und $s$ seien null- bzw.\ einstellige Konstruktoren.
\\
Gegebene Klauseln:
\\
\begin{tabular}[t]{@{}ll@{$\;$}l@{}}
1. & $p(ssx:sy)$ & $\la p(sx:y)$	\\
2. & $p(x:0)$	\\
3. & $q(sx:sx)$ & $\la q(x:x)$	\\
4. & $q(0:x)$	\\
\end{tabular}

Berechnung der Konjunktion von $p$ und $q$ nach Algorithmus \eqr{4}:
\\
\begin{tabular}[t]{@{}ll@{$\;$}ll@{}}
5. & $pq(ssx:ssx)$ & $\la pq(sx:sx)$ & aus 1.\ und 3.	\\
6. & $pq(0:0)$ && aus 2.\ und 4.	\\
\end{tabular}

Es gilt $pq(x:y) \Lra p(x:y) \wedge q(x:y)$;
die Erf\"ullbarkeit von $pq$ ist wegen 6.\ trivial.

Feststellung der Unerf\"ullbarkeit von $pq(ssx:ssx)$
nach Algorithmus \eqr{3}:
\\
\begin{tabular}[t]{@{}l@{$\;$}llll@{}}
& $inh_\vee(pq^{ssx:ssx},\{\})$	\\
$\Lra$ & $inh_\wedge(pq^{ssx:ssx},\{\})$ & Regel \eqr{3}.1	\\
$\Lra$ & $inh_\vee(pq^{sx:sx},\{pq^{ssx:ssx}\})$ 
	& Regel \eqr{3}.3	\\
$\Lra$ & $inh_\wedge(pq^{ssx:ssx},\{pq^{ssx:ssx}\})$
	& Regel \eqr{3}.1	\\
$\Lra$ & $false$ & Regel \eqr{3}.2	\\
\end{tabular}
}

Im folgenden wird eine konkrete Ordnung $\precs$ konstruiert, die die
Bedingungen aus Definition \eqr{2} erf\"ullt.
Dazu wird in Abschnitt 
\ref{Terme als Pfadmengen mit Kongruenzrelationen}
zun\"achst eine Termdarstellung ein\-ge\-f\"uhrt, die invariant
gegen gebundene Umbenennung ist und jeden Term durch seine Baumpfade und
die Kongruenzrelation gleicher Unterterme (insbesondere Variablen)
repr\"asentiert.

\subsection{Terme als Pfadmengen mit Kongruenzrelationen}
\label{Terme als Pfadmengen mit Kongruenzrelationen}

Sei ``$.$'' eine assoziative Operation mit ``$\eps$'' als neutralem
Element.
Die Menge $\P$ aller markierten Baumpfade ist definiert als kleinste
Menge mit folgenden Eigenschaften:
	\display{}{
	\begin{tabular}[t]{@{}ll@{}}
	$\eps \in \P$ 	\\
	$f \in \P$ & falls $f \in \CR$ nullstellig	\\
	$f.i.p \in \P$ & falls $f \in \CR$ $n$-stellig,
		$i \in \{1,\ldots,n\}$, $p \in \P$	\\
	\end{tabular}
	}

Die Menge $\pathps(t)$ aller (Teil-)Pfade eines Terms $t$ 
ist definiert durch:
	\display{}{
	\begin{tabular}[t]{@{}l@{$\;$}ll@{}}
	$\pathps(x)$ & $=\{\eps\}$ & f\"ur $x \in \V$	\\
	$\pathps(f)$ & $=\{f,\eps\}$ & f\"ur $f \in \CR$ nullstellig \\
	$\pathps(f(t_1,\ldots,t_n))$ 
		& $=\{f.i.p \mid p \in \pathps(t_i), 1 \leq i \leq n\} 
		\cup \{\eps\}$ 
		& f\"ur $f \in \CR$ $n$-stellig	\\
	\end{tabular}
	}

Pfade werden mit $p, p', p_i, q, \ldots$ bezeichnet,
aus dem Zusammenhang wird stets klar, ob etwa $p$ einen Pfad oder ein
Pr\"adikat meint.
F\"ur $p_1, p_2 \in \P$ definiere
	\display{}{
	\begin{tabular}[t]{@{}r@{$\;$}c@{$\;$}ll@{}}
	$p.f.i$ & $\sibl$ & $p.f.j$ 
		& f\"ur alle $p$, $f$ $n$-stellig, 
		$1 \leq i,j \leq n$ \\
	$p$ & $\pref$ & $p.p'$ & f\"ur alle $p,p'$	\\
	\mca{4}{(Gleichheit mod.\ Assoziativit\"at und Neutralit\"at)}\\
	\end{tabular}
	}

Z.B.\ f\"ur $+$ zweistellig und $s$ einstellig
ist $s.1 \pref s.1.+.1$, aber nicht $s \pref s.1.+.1$
Definiere den Unterterm $t \at p$ eines Terms $t$ am Ende eines
Pfads $p \in \pathps(t)$ durch:
	\display{}{
	\begin{tabular}[t]{@{}r@{$\;$}ll@{}}
	$t \at \eps$ & $= t$	\\
	$f(t_1,\ldots,t_n) \at {f.i.p}$ & $= t_i \at p$
		& f\"ur $1 \leq i \leq n$	\\
	\end{tabular}
	}


Zu gegebenem Term $t$ definiere $(\equiv_t)$ als die kleinste
Kongruenzrelation auf $\P$ 
mit 
$\forall p_1,p_2 \.\in P \;\;
p_1 \equiv_t p_2 \Lra t \at {p_1} = t \at {p_2}$, 
wobei rechts syntaktische Gleichheit steht (nicht mod.\ Umbenennung).
$(\equiv_t)$ mu\3 mit ``$.$'' vertr\"aglich sein in folgendem Sinne:
$p_1 \equiv_t p_2 \Ra p_1.p \equiv_t p_2.p$.
F\"ur eine beliebige Relation
$(R) \subset \P \times \P$
sei $c_{rf}(R)$ der reflexive, 
$c_{sm}(R)$ der symmetrische und
$c_{tr}(R)$ der transitive Abschlu\3 von $(R)$;
f\"ur eine Kongruenzrelation $(\equiv)$ definiere dar\"uber hinaus
die folgenden Abschlu\3operationen,
wobei $P \subset \P$ sei:
	\display{}{
	\begin{tabular}[t]{@{}l@{$\;$}ll@{}}
	$c_{mx}(\equiv)$ & $:= (\equiv) \cup 
		\{ \tpl{p_1,p_2} \mid 
			p_1.f.i \equiv p_2.f.i 
			\mbox{ f\"ur alle } i=1,\ldots,n,
			\mbox{ $f$ $n$-stellig},\}$	\\
	$c_{mx}^\infty(\equiv)$ 
		& $:= \bigcup_{i \in \N} c_{mx}^i(\equiv)$\\
	$c_{(\equiv)}(P)$
		& $:= \{ p_1 \mid p_1 \equiv p_2 \in P\}$\\
	\end{tabular}
	}

Eine Substitution $\beta$ hei\3t linear,
wenn $\beta \tpl{x_1,\ldots,x_n}$ linearer Term ist,
wobei $dom(\beta) = \{x_1,\ldots,x_n\}$ sei.
Eine Substitution $\beta$ hei\3t flach,
wenn $\beta x \in \V$ f\"ur alle $x \in dom(\beta)$.
Eine lineare und flache Substitution hei\3t Umbenennungssubstitution.
Jede Substitution $\beta$ l\"a\3t sich in eine flache und
eine lineare Substitution zerlegen:
$\beta = \beta_1 \circ \beta_2$
mit $\beta_1$ flach und $\beta_2$ linear.

\satZ{\eqd{7}}{
Zu $P \subset \P$
und $(\equiv)$ Kongruenzrelation auf $P$
gibt es einen Term $t$ mit $\pathps(t) = P$ und $(\equiv_t) = (\equiv)$
genau dann, wenn
\be
\item $\{\} \neq P \subset \P$ endlich,
\item $P$ ist abgeschlossen gegen $\sibl \circ \pref$,
	\\
	d.h.\ f\"ur $p \in P$ und $p' \sibl \circ \pref p$ 
	ist auch $p' \in P$,
\item $P$ ist abgeschlossen gegen $(\equiv)$,
	d.h.\ f\"ur $p_1 \in P$ und $p_1 \equiv p_2$
	folgt $p_2 \in P$,
\item $P$ ist vertr\"aglich mit $(\equiv)$,
	\\
	d.h.\ f\"ur $p_1 \equiv p_2$ 
	und $p_1.f_1.p'_1 \; , \; p_2.f_2.p'_2 \in P$
	folgt $f_1 = f_2$,
\item $(\equiv)$ ist Kongruenz,
	d.h.\ f\"ur $p_1 \equiv p_2$ 
	ist auch $p_1.p \equiv p_2.p$
\item $(\equiv)$ ist maximal,
	\\
	d.h.\ falls $f$ $n$-stellig 
	und $p_1.f.i \equiv p_2.f.i$ f\"ur alle $i=1,\ldots,n$,
	ist schon $p_1 \equiv p_2$,
\ee
}

Bedingung
2.\ ist \"aquivalent zur Abgeschlossenheit gegen $\sibl$ und gegen
$\pref$, da beide Relationen reflexiv sind.
Man beachte, da\3 1.\ und 2.\ automatisch erf\"ullt sind, 
wenn $P = \pathps(t')$ f\"ur ein $t'$ ist;
5.\ und 6.\ sind automatisch erf\"ullt, 
wenn $(\equiv) = (\equiv_{t'})$ f\"ur ein $t'$ ist.

In Bedingung 1.\ kann die Forderung nach Endlichkeit von $P$ auch
weggelassen werden, man erh\"alt dann die gleichen Resultate f\"ur
unendliche Terme.
Bedingung 4.\ ist ebenfalls
nicht unbedingt notwendig.
Auf sie zu verzichten hie\3e, Terme zu verallgemeinern auf gegen
Generalisierung (Anti-Instanzen) abgeschlossene Termmengen.
Damit w\"urde $lci$ zu einer totalen Operation und man erhielte einen
Verband statt eines oberen Halbverbands.
Allerdings ist unklar, ob es daf\"ur eine Anwendung gibt.

Es gelten folgende {\bf S\"atze}:
\bi
\item[\bf \eqd{8}:] Es gibt ein $\beta$ mit $t = \beta t'$
	gdw.\ $\pathps(t') \subset \pathps(t)$
	und $(\equiv_{t'}) \subset (\equiv_t)$.
	\\
	\begin{tabular}[t]{@{}ll@{}}
	In diesem Fall gilt
		& $\beta$ linear genau dann, 
		wenn $(\equiv_{t'}) = (\equiv_t)$,\\
	sowie
		& $\beta$ flach genau dann, 
		wenn $\pathps(t') = \pathps(t)$. \\
	\end{tabular}
\item[\bf \eqd{9}:] $t$ ist durch $\tpl{\pathps(t),(\equiv_t)}$
	bis auf Umbenennung eindeutig bestimmt.
\item[\bf \eqd{10}:] Sei $t = lci(t_1,\ldots,t_n)$,
	dann gilt:
	$(\equiv_t) 
	= c_{mx}^\infty \circ c_{tr}
	((\equiv_{t_1}) \cup \ldots \cup (\equiv_{t_n}))$
	und
	$\pathps(t)
	= c_{(\equiv_t)}(\pathps(t_1) \cup \ldots \cup \pathps(t_n))$.
\item[\bf \eqd{11}:] Sei $t = lci(t_1,\ldots,t_n)$, dann gilt:
	$(\equiv_{t_i}) \subset (\equiv)$ f\"ur $i=1,\ldots,n$
	$\Ra$ $(\equiv_t) \subset (\equiv)$,
	\\
	insbesondere:
	$t_i$ linear f\"ur $i=1,\ldots,n$ $\Ra$ $t$ linear.
\ei

Abbildung \ref{Beispiel zu Satz \eqr{12}}
auf Seite \pageref{Beispiel zu Satz \eqr{12}}
zeigt in der oberen und der unteren Zeile jeweils ein
Unifikationsbeispiel; punktierte Linien deuten die Kongruenzrelation
an.

\subsection{Streichungen in Pfaden}
\label{Streichungen in Pfaden}

Mit der Termdarstellung aus Abschnitt 
\ref{Terme als Pfadmengen mit Kongruenzrelationen}
kann nun eine Ordnung auf Termen definiert werden, die die
Anforderungen aus Definition \eqr{2} erf\"ullt.

Eine Streichung ist ein Paar von Pfaden,
geschrieben $q \la q.q'$.
Definiere die Ausf\"uhrung der Streichung durch:
\\
\begin{tabular}[b]{@{}ll@{$\;$}ll@{}}
1. & $del(q \la q.q',p)$ & $= \{p\}$ & falls $q \npref p$	\\
2. & $del(q \la q.q',q.p)$ & $= \{\}$ & falls $q.q' \npref q.p$	\\
3. & $del(q \la q.q',q.q'.p)$ & $= \{q.p\}$ & sonst	\\
\end{tabular}
\hfill
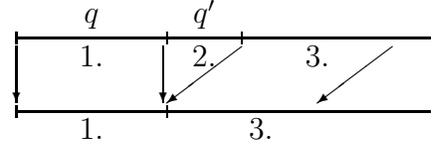
\begin{picture}(5.5,0)
\put(0.000,0.400){\line(1,0){5.500}}
	\put(0.000,0.325){\line(0,1){0.150}}
	\put(2.000,0.325){\line(0,1){0.150}}
	\put(1.000,0.300){\makebox(0.000,0.000)[t]{1.}}
	\put(3.250,0.300){\makebox(0.000,0.000)[t]{3.}}
\put(0.000,1.375){\line(1,0){5.500}}
	\put(0.000,1.300){\line(0,1){0.150}}
	\put(2.000,1.300){\line(0,1){0.150}}
	\put(3.000,1.300){\line(0,1){0.150}}
	\put(1.000,1.500){\makebox(0.000,0.000)[b]{$q$}}
	\put(2.500,1.500){\makebox(0.000,0.000)[b]{$q'$}}
	\put(1.000,1.275){\makebox(0.000,0.000)[t]{1.}}
	\put(2.500,1.275){\makebox(0.000,0.000)[t]{2.}}
	\put(4.000,1.275){\makebox(0.000,0.000)[t]{3.}}
\put(0.000,1.262){\vector(0,-1){0.750}}
\put(1.950,1.262){\vector(0,-1){0.750}}
\put(3.000,1.262){\vector(-4,-3){1.000}}
\put(5.000,1.262){\vector(-4,-3){1.000}}
\end{picture}

$del(q \la q.q',p)$ liefert eine maximal einelementige Pfadmenge.
F\"ur eine Pfadmenge $P$ definiere
$del(q \la q.q',P) := \bigcup_{p \in P} del(q \la q.q',p)$;
f\"ur eine Folge von Streichungen 
$S = \tpl{q_1 \la q_1.q'_1, \ldots, q_n \la q_n.q'_n}$
definiere 
$del(S,P) 
:= del(q_n \la q_n.q'_n, \ldots del(q_1 \la q_1.q'_1,P) \ldots )$.
Es gilt stets $del(S,p_1) = del(S,p_2) \neq \{\} \;\Ra\; p_1 = p_2$.

Eine Streichungsfolge $S$ auf $P \subset \P$
hei\3t vertr\"aglich mit einer Kongruenzrelation $(\equiv)$
genau dann, wenn 
$del(S,p_1) \neq \{\} \Lra del(S,p_2) \neq \{\}$
f\"ur alle $p_1 \equiv p_2$
ist.
In diesem Fall definiere 
$p'_1 \equiv_S p'_2 :\Lra p_1 \equiv p_2$
f\"ur $del(S,p_1) = \{p'_1\}$ und $del(S,p_2) = \{p'_2\}$,
sowie
$del(S,(\equiv)) := c_{mx}^\infty(\equiv_S)$.

Ist $S$ vertr\"aglich mit $(\equiv_t)$,
so gibt es genau ein $t'$ mit 
\bi
\item $\pathps(t') = del(S,\pathps(t))$,
\item $(\equiv_{t'}) = del(S,(\equiv_t))$ und
\item $t' \at {p'} = t \at p$
	f\"ur alle $p \in \pathps(t)$ 
	mit $t \at p \in \V$ 
	und $del(S,p) = \{p'\}$.
\ei

Definiere $del(S,t) := t'$
sowie $t' \preceq t :\Lra \exists S \;\; t' = del(S,t)$.

Die Wohlfundiertheit von $\prec$ folgt daraus, da\3 $\prec$ eine
Unterrelation der Einbettungsordnung nach
\cit{Huet86} ist.
Auf der Menge der linearen Terme sind beide Ordnungen identisch.
Abbildung \ref{Beispiel zu Satz \eqr{12}}
zeigt in der linken, mittleren und rechten Spalte jeweils eine
Anwendung von $del(f.3 \la f.3.g_3.1,\ldots)$.

\begin{figure}
\begin{center}
\begin{picture}(12,9)
\put(1.000,3.000){\makebox(0.000,0.000){$f$}}
\put(0.000,2.000){\makebox(0.000,0.000){$g_1$}}
\put(1.000,2.000){\makebox(0.000,0.000){$x_2$}}
\put(2.000,2.000){\makebox(0.000,0.000){$x_2$}}
\put(0.000,1.000){\makebox(0.000,0.000){$x_1$}}
\put(0.800,2.800){\line(-1,-1){0.600}}
\put(1.000,2.800){\line(0,-1){0.600}}
\put(1.200,2.800){\line(1,-1){0.600}}
\put(0.000,1.800){\line(0,-1){0.600}}
\multiput(1.200,2.000)(0.070,0.000){8}{$\scriptscriptstyle \cdot$}
\put(6.000,3.000){\makebox(0.000,0.000){$f$}}
\put(5.000,2.000){\makebox(0.000,0.000){$g_1$}}
\put(6.000,2.000){\makebox(0.000,0.000){$g_2$}}
\put(7.000,2.000){\makebox(0.000,0.000){$g_2$}}
\put(5.000,1.000){\makebox(0.000,0.000){$x_1$}}
\put(6.000,1.000){\makebox(0.000,0.000){$g_1$}}
\put(7.000,1.000){\makebox(0.000,0.000){$g_1$}}
\put(6.000,0.000){\makebox(0.000,0.000){$x_1$}}
\put(7.000,0.000){\makebox(0.000,0.000){$x_1$}}
\put(5.800,2.800){\line(-1,-1){0.600}}
\put(6.000,2.800){\line(0,-1){0.600}}
\put(6.200,2.800){\line(1,-1){0.600}}
\put(5.000,1.800){\line(0,-1){0.600}}
\put(6.000,1.800){\line(0,-1){0.600}}
\put(7.000,1.800){\line(0,-1){0.600}}
\put(6.000,0.800){\line(0,-1){0.600}}
\put(7.000,0.800){\line(0,-1){0.600}}
\multiput(5.200,1.800)(0.050,-0.050){12}{$\scriptscriptstyle \cdot$}
\multiput(6.200,2.000)(0.070,0.000){8}{$\scriptscriptstyle \cdot$}
\multiput(5.200,0.800)(0.050,-0.050){12}{$\scriptscriptstyle \cdot$}
\multiput(6.200,1.000)(0.070,0.000){8}{$\scriptscriptstyle \cdot$}
\multiput(6.200,0.000)(0.070,0.000){8}{$\scriptscriptstyle \cdot$}
\put(11.000,3.000){\makebox(0.000,0.000){$f$}}
\put(10.000,2.000){\makebox(0.000,0.000){$x_3$}}
\put(11.000,2.000){\makebox(0.000,0.000){$g_2$}}
\put(12.000,2.000){\makebox(0.000,0.000){$x_4$}}
\put(11.000,1.000){\makebox(0.000,0.000){$x_3$}}
\put(10.800,2.800){\line(-1,-1){0.600}}
\put(11.000,2.800){\line(0,-1){0.600}}
\put(11.200,2.800){\line(1,-1){0.600}}
\put(11.000,1.800){\line(0,-1){0.600}}
\multiput(10.200,1.800)(0.050,-0.050){12}{$\scriptscriptstyle \cdot$}
\put(1.000,9.000){\makebox(0.000,0.000){$f$}}
\put(0.000,8.000){\makebox(0.000,0.000){$g_1$}}
\put(1.000,8.000){\makebox(0.000,0.000){$x_2$}}
\put(2.000,8.000){\makebox(0.000,0.000){$g_3$}}
\put(0.000,7.000){\makebox(0.000,0.000){$x_1$}}
\put(2.000,7.000){\makebox(0.000,0.000){$x_2$}}
\put(0.800,8.800){\line(-1,-1){0.600}}
\put(1.000,8.800){\line(0,-1){0.600}}
\put(1.200,8.800){\line(1,-1){0.600}}
\put(0.000,7.800){\line(0,-1){0.600}}
\put(2.000,7.800){\line(0,-1){0.600}}
\multiput(1.200,7.800)(0.050,-0.050){12}{$\scriptscriptstyle \cdot$}
\put(6.000,9.000){\makebox(0.000,0.000){$f$}}
\put(5.000,8.000){\makebox(0.000,0.000){$g_1$}}
\put(6.000,8.000){\makebox(0.000,0.000){$g_2$}}
\put(7.000,8.000){\makebox(0.000,0.000){$g_3$}}
\put(5.000,7.000){\makebox(0.000,0.000){$x_1$}}
\put(6.000,7.000){\makebox(0.000,0.000){$g_1$}}
\put(7.000,7.000){\makebox(0.000,0.000){$g_2$}}
\put(6.000,6.000){\makebox(0.000,0.000){$x_1$}}
\put(7.000,6.000){\makebox(0.000,0.000){$g_1$}}
\put(7.000,5.000){\makebox(0.000,0.000){$x_1$}}
\put(5.800,8.800){\line(-1,-1){0.600}}
\put(6.000,8.800){\line(0,-1){0.600}}
\put(6.200,8.800){\line(1,-1){0.600}}
\put(5.000,7.800){\line(0,-1){0.600}}
\put(6.000,7.800){\line(0,-1){0.600}}
\put(7.000,7.800){\line(0,-1){0.600}}
\put(6.000,6.800){\line(0,-1){0.600}}
\put(7.000,6.800){\line(0,-1){0.600}}
\put(7.000,5.800){\line(0,-1){0.600}}
\multiput(5.200,7.800)(0.050,-0.050){12}{$\scriptscriptstyle \cdot$}
\multiput(6.200,7.800)(0.050,-0.050){12}{$\scriptscriptstyle \cdot$}
\multiput(5.200,6.800)(0.050,-0.050){12}{$\scriptscriptstyle \cdot$}
\multiput(6.200,6.800)(0.050,-0.050){12}{$\scriptscriptstyle \cdot$}
\multiput(6.200,5.800)(0.050,-0.050){12}{$\scriptscriptstyle \cdot$}
\put(11.000,9.000){\makebox(0.000,0.000){$f$}}
\put(10.000,8.000){\makebox(0.000,0.000){$x_3$}}
\put(11.000,8.000){\makebox(0.000,0.000){$g_2$}}
\put(12.000,8.000){\makebox(0.000,0.000){$x_4$}}
\put(11.000,7.000){\makebox(0.000,0.000){$x_3$}}
\put(10.800,8.800){\line(-1,-1){0.600}}
\put(11.000,8.800){\line(0,-1){0.600}}
\put(11.200,8.800){\line(1,-1){0.600}}
\put(11.000,7.800){\line(0,-1){0.600}}
\multiput(10.200,7.800)(0.050,-0.050){12}{$\scriptscriptstyle \cdot$}
\thicklines
\put(1.000,4.500){\vector(0,-1){1.000}}
	\put(1.100,4.000){\makebox(0,0)[l]{$\scriptstyle less$}}
\put(6.000,4.500){\vector(0,-1){1.000}}
	\put(6.100,4.000){\makebox(0,0)[l]{$\scriptstyle less$}}
\put(11.000,4.500){\vector(0,-1){1.000}}
	\put(11.100,4.000){\makebox(0,0)[l]{$\scriptstyle less$}}
\put(3.000,2.000){\vector(1,0){1.000}}
	\put(3.50,2.100){\makebox(0,0)[b]{$\scriptstyle lci$}}
\put(9.000,2.000){\vector(-1,0){1.000}}
	\put(8.50,2.100){\makebox(0,0)[b]{$\scriptstyle lci$}}
\put(3.000,8.000){\vector(1,0){1.000}}
	\put(3.50,8.100){\makebox(0,0)[b]{$\scriptstyle lci$}}
\put(9.000,8.000){\vector(-1,0){1.000}}
	\put(8.50,8.100){\makebox(0,0)[b]{$\scriptstyle lci$}}
\end{picture}
\caption{Beispiel zu Satz \eqr{12}}
\label{Beispiel zu Satz \eqr{12}}
\end{center}
\end{figure}

\satZ{\eqd{12}}{\\
Zu $t' \preceq t = lci(t_1,\ldots,t_n)$
existieren $t'_1 \preceq t_1, \ldots, t'_n \preceq t_n$
mit $t' = lci(t'_1,\ldots,t'_n)$.
\\
Siehe Beispiel in Abbildung \ref{Beispiel zu Satz \eqr{12}}.
}

F\"ur $t' = del(S,t)$ ist nicht immer $\beta t' = del(S,\beta t)$,
sondern nur dann, wenn $S$ nicht nur mit $(\equiv_t)$, sondern
dar\"uber hinaus auch mit $(\equiv_{\beta t})$ vertr\"aglich ist.
Zur Sicherstellung der Forderung \eqr{2}.5 verlangen wir daher die
Existenz einer globalen Kongruenzrelation $(\equivs)$ mit folgenden
Eigenschaften:

\dfn{\eqd{13}}{
\be
\item $(\equivs)$ ist maximal im Sinne von \eqr{7}.6,
	d.h.\ abgeschlossen gegen $c_{mx}$,
\item $(\equivs)$ ist abgeschlossen gegen konsistente Streichungen,
	\\
	d.h.\ f\"ur $(\equiv_t) \subset (\equivs)$
	und $S$ vertr\"aglich mit $(\equivs)$
	ist auch $(\equiv_{del(S,t)}) \subset (\equivs)$.
\ee
}

Von einem wie in \eqr{1} gegebenen Horn-Programm
verlangen wir, da\3 $(\equiv_\tpl{t_i,t_{i'}}) \subset (\equivs)$
und $(\equiv_{t_i}) \subset (\equivs)$
f\"ur alle $1 \leq i,i' \leq m$.
Man beachte, da\3 $(\equivs)$ i.allg.\ nicht als
$(\equiv_t)$ f\"ur irgend einen Term $t$ darstellen l\"a\3t.

\dfn{\eqd{14}}{
Wir definieren $t' \preceqs t$ genau dann, 
wenn $(\equiv_t) \subset (\equivs)$ ist und es eine mit $(\equivs)$
vertr\"agliche Streichungsfolge $S$ gibt, so da\3 $t' = del(S,t)$.
}

Als Unterrelation von $\prec$ ist $\precs$ ebenfalls Wohlordnung, und
Satz \eqr{12} folgt leicht auch f\"ur $\preceqs$.
Damit erf\"ullt $\preceqs$ alle Forderungen aus Definition \eqr{2},
wobei \eqr{2}.5 pr\"azisiert wird zu
	\display{\eqr{2}.5'}{
	$t' \preceqs t \;\wedge\; (\equiv_{\beta t}) \subset (\equivs) 
	\;\Ra\; \beta t' \preceqs \beta t$.
	}

Die dadurch in den Terminierungsbeweisen zus\"atzlich notwendigen
Voraussetzungen kann man mit Hilfe von Satz \eqr{11} leicht
zeigen;
etwa in der zweiten Zeile von Fall 3.\ im Beweis von Lemma \eqr{5}
ist $(\equiv_{\beta_i t_i}) \subset (\equivs)$,
da $(\equiv_{t_i}) \subset (\equivs)$
und $(\equiv_{t'}) \subset (\equivs)$ nach Induktionsvoraussetzung,
damit ist nach \eqr{13}.2 
auch $(\equiv_{\beta_i t_{i1}}) \subset (\equivs)$.

In Beispiel \eqr{6} wird $(\equivs)$ induziert von
$(:.1) \equivs (:.2)$,
und die einzige vorkommende Streichungsfolge ist
$\tpl{(:.1) \la (:.1.s.1) , (:.2) \la (:.2.s.1)}$.
An einer hinreichenden Bedingung f\"ur \eqr{13}.2, die die konstruktive
Berechnung
von $(\equivs)$ aus den in einem gegebenen Horn-Programm
vorkommenden $(\equiv_{t_i})$ und $(\equiv_{t_{ij}})$ erlaubt,
wird z.Zt.\ gearbeitet.

\subsection{Ausblick}
\label{Ausblick}

Dar\"uber hinaus scheint es interessant, die sich aus Definition
\eqr{2}
ergebenden Forderungen an $\preceqs$ zun\"achst abstrakt zu formulieren
und weitere konkrete ``Implementierungen'' von $\preceqs$ zu
untersuchen.
In diesem Zusammenhang werden Abbildungen
$\phi: \T \ra \T$ untersucht, f\"ur die gilt
	\display{}{
	\begin{tabular}[t]{@{}r@{$\;$}ll@{}}
	$\phi(lci(t_1,t_2))$ & $= lci(\phi(t_1),\phi(t_2))$ & und \\
	$\phi(t)$ & $\leq t$ 
		& bzgl.\ einer beliebigen Wohlordnung $<$	\\
	\end{tabular}
	}

Eine solche Abbildung ist durch ihr Verhalten auf $\T_1 \cup \T_2$
schon eindeutig bestimmt, wobei
	\display{}{
	\begin{tabular}[t]{@{}l@{$\;$}lll@{}}
	$\T_1$ & $:= \{ t \in \T \mid \exists p \in \P$
		& $\pathps(t) = c_{\sibl \circ \pref}(\{p\}) \}$
		& und \\
	$\T_2$ & $:= \{ t \in \T \mid \exists p_1 \neq p_2 \in \P$
		& $\pathps(t) = c_{\sibl \circ \pref}(\{p_1,p_2\}),$ \\
		&& $(\equiv_t) 
		= c_{sm} \circ c_{rf} (\{\tpl{p_1,p_2}\}) \}$ & .\\
	\end{tabular}
	}

Auf Seite \pageref{T1T2}
sind zwei Elemente von $\T_1$ bzw.\ $\T_2$ skizziert.
$lci(\T_1)$ ist gerade die Menge aller linearen Terme;
und $lci(\T_1 \cup \T_2) = \T$.
F\"ur $lci(t_1,t_2) = t \in \T_1 \cup \T_2$ 
mu\3 schon $t = t_1$ oder $t = t_2$ bis auf Umbenennung sein.
Es gibt eine notwendige und hinreichende Bedingung daf\"ur, da\3 sich
eine vorgegebene Zuordnung
$t_1 \mapsto t_{11}, \ldots, t_m \mapsto t_{m1}$
zu einer Abbildung $\phi$ mit den obigen Eigenschaften fortsetzen
l\"a\3t.

\parbox[b]{11.0cm}{
Damit l\"a\3t sich $\preceqs$ allgemeiner als in Def.\ \eqr{14}
definieren durch $t' \preceqs t :\Lra \exists \phi \;\; t' = \phi(t)$;
und es k\"onnen
auch noch Horn-Programme wie das f\"ur $reverse$ behandelt werden,
in denen beim \"Ubergang vom Kopf zum Rumpf
Konstruktorsymbole von einem Argument in ein anderes (lexikographisch
nachrangiges) geschoben werden.
}
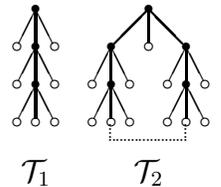
\hfill
\begin{picture}(2.5,0)
\label{T1T2}
\put(0.250,0.500){\makebox(0.000,0.000)[t]{$\T_1$}}
\put(0.250,2.500){\circle*{0.100}}
	\put(0.250,2.500){\line(-1,-2){0.225}}
	{\thicklines \put(0.250,2.500){\line(0,-1){0.450}}}
	\put(0.250,2.500){\line(1,-2){0.225}}
\put(0.000,2.000){\circle{0.100}}
\put(0.250,2.000){\circle*{0.100}}
\put(0.500,2.000){\circle{0.100}}
	\put(0.250,2.000){\line(-1,-2){0.225}}
	{\thicklines \put(0.250,2.000){\line(0,-1){0.450}}}
	\put(0.250,2.000){\line(1,-2){0.225}}
\put(0.000,1.500){\circle{0.100}}
\put(0.250,1.500){\circle*{0.100}}
\put(0.500,1.500){\circle{0.100}}
	\put(0.250,1.500){\line(-1,-2){0.225}}
	{\thicklines \put(0.250,1.500){\line(0,-1){0.450}}}
	\put(0.250,1.500){\line(1,-2){0.225}}
\put(0.000,1.000){\circle{0.100}}
\put(0.250,1.000){\circle{0.100}}
\put(0.500,1.000){\circle{0.100}}

\put(1.750,0.500){\makebox(0.000,0.000)[t]{$\T_2$}}
\put(1.750,2.500){\circle*{0.100}}
	{\thicklines \put(1.750,2.500){\line(-1,-1){0.450}}}
	\put(1.750,2.500){\line(0,-1){0.450}}
	{\thicklines \put(1.750,2.500){\line(1,-1){0.450}}}
\put(1.250,2.000){\circle*{0.100}}
\put(1.750,2.000){\circle{0.100}}
\put(2.250,2.000){\circle*{0.100}}
	\put(1.250,2.000){\line(-1,-2){0.225}}
	{\thicklines \put(1.250,2.000){\line(0,-1){0.450}}}
	\put(1.250,2.000){\line(1,-2){0.225}}
\put(1.000,1.500){\circle{0.100}}
\put(1.250,1.500){\circle*{0.100}}
\put(1.500,1.500){\circle{0.100}}
	\put(1.250,1.500){\line(-1,-2){0.225}}
	{\thicklines \put(1.250,1.500){\line(0,-1){0.450}}}
	\put(1.250,1.500){\line(1,-2){0.225}}
\put(1.000,1.000){\circle{0.100}}
\put(1.250,1.000){\circle{0.100}}
\put(1.500,1.000){\circle{0.100}}
	\put(2.250,2.000){\line(-1,-2){0.225}}
	{\thicklines \put(2.250,2.000){\line(0,-1){0.450}}}
	\put(2.250,2.000){\line(1,-2){0.225}}
\put(2.000,1.500){\circle{0.100}}
\put(2.250,1.500){\circle*{0.100}}
\put(2.500,1.500){\circle{0.100}}
	\put(2.250,1.500){\line(-1,-2){0.225}}
	{\thicklines \put(2.250,1.500){\line(0,-1){0.450}}}
	\put(2.250,1.500){\line(1,-2){0.225}}
\put(2.000,1.000){\circle{0.100}}
\put(2.250,1.000){\circle{0.100}}
\put(2.500,1.000){\circle{0.100}}
\multiput(1.200,0.700)(0.050,0.000){21}{$\scriptscriptstyle\cdot$}
\multiput(1.200,0.900)(0.000,-0.050){4}{$\scriptscriptstyle\cdot$}
\multiput(2.200,0.900)(0.000,-0.050){4}{$\scriptscriptstyle\cdot$}
\end{picture}

\subsection{Literatur}

\bd


\dt{Bogaert92} Equality and disequality constraints on direct subterms
	in tree automata, B.\ Bogaert, S.\ Tison, in: Proc.\ Symposium
	on Theoretical Aspects of Computer Science, Springer, P.\
	161-171, 1992

\dt{Burghardt93} Eine feink\"ornige Sortendisziplin und ihre Anwendung
	in der Programmkonstruktion, Jochen Burghardt, Dissertation,
	GMD Report 212, Oldenbourg, 1993

\dt{Burghardt95} A decidable class of $n$-ary Horn predicates, J.\
	Burghardt, Technical Report, GMD, forthcoming

\dt{Comon90} Equational formulas in order-sorted algebras, Hubert
	Comon, in: Proc.\ ICALP, Warwick, Springer-Verlag, Jul 1990

\dt{Comon92} On unification of terms with integer exponents,
	Hubert Comon, Research Report 770, L.R.I., Univ. Paris-Sud,
	Orsay, 1992

\dt{Heinz94} Lemma Discovery by Anti-Unification of Regular Sorts,
	Birgit Heinz, TU Berlin, Technical Report 94-21, 1994

\dt{Huet86} Computation and Deduction, Huet, G.,
	International Summer School on Logic of Programming and
	Calculi of Discrete Design, (Chapter 4: Termination),
	Marktoberdorf 1986

\dt{Jones79}
	Flow analysis and optimization of LISP-like structures, Jones,
	N.D., Muchnick, S.S., in: Sixth Annual ACM Symposium on
	Principles of Programming Languages, p.~244--256, Jan 1979

\dt{Mishra84} Towards a theory of types in Prolog,
	P.\ Mishra, in: Proc.\ 1984 Inter.\ Symp.\ on Logic Programming,
	p.\ 289-298, 1984

\dt{Schmidt-Schauss88} Computational aspects of an order-sorted
	logic with term declarations, M.\ Schmidt-Schau\3 , Univ.\
	Kaiserslautern, Dissertation, FB Informatik, April 1988,
	LNAI 395, 1989

\dt{Tommasi91} Automates avec tests d'\'egalit\'es entre cousins
	germains, M.\ Tommasi, LIFL-IT report, 1991

\dt{Uribe92} Sorted Unification Using Set Constraints,
	T.E.\ Uribe, in: Proc.\ CADE-11, LNCS 607, p.\ 163-177, 1992

\ed

\end{document}